\documentstyle[prc,aps,preprint,psfig]{revtex}
\draft
\begin{document}
\title{$e^+e^-$ pairs from $\pi^-$A reactions
\thanks{Work supported by BMBF and GSI Darmstadt.}}
\author{Th. Weidmann, E. L. Bratkovskaya, W. Cassing, and U. Mosel \\[5mm]
Institut f\"ur Theoretische Physik,
Universit\"at Giessen\\D-35392 Giessen, Germany }
\maketitle

\begin{abstract}
We present a dynamical study of $e^+e^-$ production in  $\pi^-$A reactions
at 1.3 and 1.7 GeV on the basis of the Coupled-Channel-BUU approach.
The contributions from vector mesons ($\rho,\omega$) are calculated
taking into account the collisional broadening effect and are compared
to background sources in the dilepton spectrum from
the Dalitz decays of $\omega$ and $\eta$
mesons produced in the reaction. Two possible scenarios for the medium
modifications of the vector mesons are investigated, i.e.
the 'dropping mass' scheme for the $\rho$ and $\omega$ and a momentum
dependent $\rho$-meson spectral function that includes the polarization
of the $\rho$-meson due to resonant $\rho -N$ scattering.
\end{abstract}

\vspace*{1cm}


\narrowtext
\newpage
\section{Introduction}

The properties of hadrons in the nuclear medium are of fundamental interest
(cf. Refs. \cite{BrownRho,Shakin94,Klingl96,H&L92,Asakawa93}).
According to QCD sum rules \cite{H&L92,Asakawa93,Leupold} or QCD
inspired  effective Lagrangian models
\cite{BrownRho,Shakin94,Klingl96,Herrmann,asakawa,Chanfray,Rapp,Friman,RappNPA,Peters}
the properties of the vector mesons ($\rho$, $\omega$ and $\phi$)
should change with the nuclear density. Furthermore, along with a
dropping mass the phase space for the resonance decay also decreases
which results in a modification of the resonance lifetime in matter.  On
the other hand, due to collisional broadening -- which depends on the
nuclear density and the resonance-nucleon interaction cross section (
cf. Refs.  \cite{Kondr94,Boresk96}) -- the resonance lifetime decreases
again.

The in-medium properties of vector mesons have been studied
experimentally so far by dilepton measurements at SPS energies for
proton-nucleus and nucleus-nucleus collisions
\cite{CERES,Ullrich,HELIOS,HELI2}.  As proposed by Li, Ko, and Brown
\cite{Li} and Ko et al. \cite{Li96}, the observed enhancement in
A + A reactions compared to p + A collisions in the invariant mass
range $0.3 \leq M \leq 0.7$ GeV might be due to a shift of the
$\rho$-meson mass following Brown/Rho scaling \cite{BrownRho} or the
Hatsuda and Lee sum rule prediction~\cite{H&L92}.  The microscopic
transport studies in Refs. \cite{Cass95C,Cass96H,Brat97} for these
systems support these results \cite{Li,Li96}.

However, also a more conventional approach including the change of the
$\rho$-meson spectral function in the medium due to the coupling of
$\rho, \pi, \Delta$ and nucleon dynamics along the lines of
Refs.~\cite{Herrmann,asakawa,Chanfray,Rapp} was found to be (roughly)
compatible with the CERES data~\cite{Rapp,Cass95C}. Meanwhile, our
knowledge on the $\rho$ spectral function has improved since -- as
first pointed out by Friman and Pirner~\cite{Friman} -- resonant
$\rho$-$N$ interactions significantly enhance the strength in the
vector-isovector channel at low invariant mass; this has also been
confirmed recently in Ref.  \cite{Peters} where the authors find a
significant smearing of the $\rho$ strength at low $\rho$ momenta due
to a selfconsistent evaluation of the resonance widths.  In fact, the
CERES data for S~+~Au at 200~A~GeV and Pb~+~Au at 160~A~GeV, using an
expanding fireball model, were found to be compatible with such a
hadronic scenario~\cite{RappNPA}.

Recently, the thermodynamical result of Ref.~\cite{RappNPA} was
supported by Hadron String Dynamics (HSD) transport calculations
\cite{CBRW97} where an improved $\rho$ spectral function was used.
The conclusion in Ref.~\cite{CBRW97} was  that both the 'dropping mass'
scenario~\cite{Li96,Cass95C,Cass96H,Brat97} as well as the hadronic
spectral function approach~\cite{RappNPA} lead to dilepton spectra
which are in good agreement with the experimental data for all systems
at SPS energies.  A similar analysis was done also for BEVALAC/SIS
energies \cite{BCRW97} (where quite different temperature and density
regimes are probed) involving both the spectral functions from
Refs.~\cite{RappNPA,Peters}.  It was found that these spectral
functions give practically the same result for dilepton spectra at
BEVALAC/SIS energies.

Additional, and possibly more sensitive, information could be provided by
studies using more elementary probes such a pions or protons as incoming
particles. In such reactions the nuclear density ($\le \rho_0$) is
obviously lower than in heavy-ion collisions, but the phase-space
distribution of the  nuclear matter is almost stationary and much better
known.  In case of pion-nucleus reactions the $\omega$ meson can be
produced with low momenta in the laboratory system such that a
substantial fraction of them will also decay inside a heavy nucleus
\cite{Schoen,CGIK97,GKC97}.  In Refs.~\cite{CGIK97,GKC97}, which employed
the Hatsuda-Lee mass shift \cite{H&L92}, it was shown
on the basis of the intranuclear cascade (INC) approach that the mass
distributions of the vector mesons decaying inside the nucleus have a
two-component structure \cite{Boresk96} in the dilepton invariant  mass
spectrum: the first, high-mass component corresponds to resonances
decaying in the vacuum, thus showing the free spectral function which
is very narrow in case of the $\omega$ meson; the second (broader)
component of lower masses corresponds to the resonance decay inside the
nucleus.

The many detailed hadronic model studies quoted above have shown that
the assumption of a very narrow, unchanged width of the vector mesons,
that underlies the Hatsuda-Lee prescription, is oversimplified.
This has also recently been discussed in Ref.~\cite{Leupold}, where
the authors have shown that QCD sum rules pose only very loose constraints
on the in-medium properties of vector mesons.

In this paper we, therefore, report results of detailed microscopic
calculations on the basis of the
Coupled-Channel-Boltzmann-Uehling-Uhlenbeck (CBUU) model \cite{TeisZP97}
for the dilepton production in $\pi^-$C, $\pi^-$Ca,  $\pi^-$Pb collisions
at kinetic energies $E_{kin}=1.3$ and 1.7~GeV employing also a realistic
spectral function for the $\rho$ meson \cite{Peters} in addition to the
collisional broadening and 'dropping' mass effects of vector mesons
studied in Refs.~\cite{CGIK97,GKC97}.  Within this transport model it
is possible to investigate simultaneously $\gamma$ + A, $\pi$ + A, p +
A and A + A reactions in a wide dynamical range such that especially
the pion dynamics can be controlled by a large set of independent
experiments \cite{TeisZP97}.

Our paper is organized as follows: In Section 2 we briefly describe the
CBUU transport approach and its implementation for pion-induced
reactions, the $\pi N$ total reaction cross section and the elementary
processes for meson production and their interactions employed in the
CBUU approach. In Section 3 we discuss the elementary channels for
dilepton production and collisional broadening and the in-medium
modification schemes ('dropping' mass and $\rho$ spectral function
approach) which will be applied for calculating the dilepton spectra.
Section 4 contains a detailed study of dilepton spectra  for $\pi^-$C,
$\pi^-$Ca, $\pi^-$Pb  reactions at $E_{kin}=1.3$ and 1.7~GeV. We close
with a summary in Section 5.

\section{Ingredients of the CBUU-Model}

\subsection{Binding energy and nuclear stability}

The dynamical description of pion-nucleus collisions is performed
within the Coupled-Channel-BUU approach~\cite{TeisZP97}
which has been found to describe reasonably well various pion data at SIS
energies.  The model has been described in detail in ~\cite{TeisZP97};
here we briefly recall the main ingredients.

In line with Refs. \cite{bertsch88,cassing88,cassing90,kweber93}
the dynamical evolution of heavy-ion collisions or
hadron-nucleus reactions below the pion-production threshold is described
by a transport equation for the nucleon one-body
phase-space distribution function $f_N({\bf r},{\bf p},t)$,
\begin{eqnarray}
\label{buueq}
\frac{\partial f({\bf r},\,{\bf p},\,t)}{\partial t} &+& \left\{ \frac{{\bf p}}{E} +
\frac{m^*({\bf r},{\bf p})}{E}\, {\bf \nabla}_p\, U({\bf r},\, {\bf p}) \right\}
\, {\bf \nabla}_r f({\bf r},\,{\bf p},\,t)  \nonumber  \\
& &+ \left\{ -\frac{m^*({\bf r},{\bf p})}{E}\, {\bf \nabla}_r
U({\bf r},\, {\bf p}) \right\}\, {\bf \nabla}_p f({\bf r},\,{\bf p},\,t)
\,\, =\, \, I_{coll}[f({\bf r},\,{\bf p},\,t)],
\end{eqnarray}
where ${\bf r}$ and ${\bf p}$ denote the spatial and the momentum coordinate of
the nucleon,
respectively, while $N$ stands for a proton ($p$) or neutron ($n$).
The effective mass $m^*({\bf r}, \, {\bf p})$ in
Eq. (\ref{buueq}) includes the nucleon restmass $m_N$ ($ = 938$ MeV/$c^2$)
as well as a scalar momentum-dependent mean-field potential
$U({\bf r},\, {\bf p})$,
\begin{eqnarray}
\label{effmass}
m^*({\bf r}, \, {\bf p})  = m_N + U({\bf r},\, {\bf p})
\end{eqnarray}
for the baryons. The nucleon quasi-particle dispersion relation then is
\begin{eqnarray}
\label{esingle}
E ({\bf r},{\bf p}) = \sqrt{ m^*({\bf r}, \, {\bf p})^2 + {\bf p}^2}.
\end{eqnarray}
We note that we have neglected an explicit vector interaction  due
to numerical reasons in order to achieve a better stability of nuclei
in their 'groundstate'. However, this is of no concern since the
vector interaction plays no role in $\pi$ + A reactions as considered
in this work.

Here we employ the momentum-dependent mean-field potential proposed by
Welke et al.  \cite{welke88} with an additional isospin symmetry
potential $U^{symm}$, i.e.
\begin{eqnarray}
U^{nr}({\bf r}, \, {\bf p}) = A \frac{\rho}{\rho_0} +
B \frac{\rho}{\rho_0}^\tau + 2 \frac{C}{\rho_0}
\int d^3 p'\frac{f({\bf r},{\bf p}^\prime)}
{1 + \left(\frac{ {\bf p}-{\bf p}^\prime}{\Lambda}\right)^2} + U^{symm},
\label{welkepot}
\end{eqnarray}
where $U^{symm}$ is
\begin{eqnarray}
U^{symm}=D {\rho_p({\bf r})-\rho_n({\bf r}) \over \rho_0} \tau_z,
\label{potsymm}
\end{eqnarray}
with $\tau_z=-1$ for neutrons and $\tau_z=+1$ for protons; for the
strength of the symmetry potential we use $D=30$~MeV in line with
\cite{LiRen}.

As an extension of the momentum-independent Skyrme type
potentials for nuclear matter \cite{cassing90,welke88} the parametrization
(\ref{welkepot}) has no manifest Lorentz-properties.
However, definite Lorentz-properties are required for a transport model at
relativistic energies. To achieve this goal
we evaluate the non-relativistic mean-field potential $U^{nr}$ in the
local rest frame (LRF) of nuclear matter which is defined by the frame
of reference with vanishing local vector baryon current
(${\bf j}(r,t) = {\bf 0}$).
Assuming only scalar potentials in the LRF we then equate the
expressions for the single-particle energies using the
non-relativistic potential $U^{nr}$ and the scalar potential $U$ by
\begin{eqnarray}
\sqrt{p^2 + m^2} + U^{nr}({\bf r},\,{\bf p}) =  \sqrt{p^2 + \left(
m + U \left({\bf r},\,{\bf p} \right) \right)^2}.
\label{defuscal}
\end{eqnarray}
Eq. (\ref{defuscal}) now allows to extract the scalar mean-field potential
$U({\bf r},\,{\bf p})$ which we will use throughout our
calculations for the baryons.

For our calculations we use a (momentum-dependent) equation of
state (EOS) for nuclear matter with an incompressibility of $K = 290$~MeV
(i.e. $A = -29.3$~MeV, $B = 57.2$~MeV, $C = -63.5$~MeV, $\tau$ = 1.760,
$\Lambda$ = 2.13 fm$^{-1}$). For the pion-nucleus reactions to be
investigated in this study, however, the nuclear incompressibility $K$
is no decisive quantity and is only quoted for completeness.

For $\pi$ + A reactions it is important
that the nuclear groundstate properties, i.e. density and binding
energies, are well reproduced. We illustrate the quality of our groundstate
in Fig.~\ref{piAfig1}, where we show the binding energy per
nucleon as a function of the nucleon number $A$.  The solid curve
corresponds to the binding energy  calculated with $U^{symm}$ whereas
the dashed curve is the result without the isospin symmetry potential;
the solid curve belongs to nuclei that are quite stable over time
periods longer than the typical reaction time of 15 fm/c in case of a
Pb-target.  Comparing our calculations for stationary nuclei with the
empirical liquid drop result (dotted line) we find that our dynamical
calculations with $U^{symm}$ agree reasonably well with the
experimental systematics.

In the CBUU-model we explicitly propagate the mesonic degrees of
freedom $\pi$, $\eta$, $\rho$ and a scalar meson $\sigma$ that
simulates correlated $2\pi$ pairs in the isospin $0$-channel.  Besides
the nucleon and the $\Delta(1232)$ we, furthermore, include all
baryonic resonances up to a mass of $1950$ MeV/c$^2$: i.e. $N(1440)$,
$N(1520)$, $N(1535)$, $\Delta(1600)$, $\Delta(1620)$, $N(1650)$,
$\Delta(1675)$, $N(1680)$, $\Delta(1700)$, $N(1720)$, $\Delta(1905)$,
$\Delta(1910)$ and $\Delta(1950)$; the resonance properties are
adopted from the PDG \cite{pdg}.

Denoting the nucleon by $N$ and the baryon resonances listed above by $R$ and
$R'$, we include  the following channels:
\begin{itemize}
\item elastic baryon-baryon collisions
$N N  \leftrightarrow  N N, N R  \leftrightarrow  N R $;
\item inelastic baryon-baryon collisions
$N N  \leftrightarrow  N R,
N R  \leftrightarrow  N R',
N N  \leftrightarrow  \Delta(1232) \Delta(1232)$;
\item inelastic baryon-meson reactions
$R  \leftrightarrow  N \pi,$ \
$R  \leftrightarrow  N \pi \pi,
   \Delta(1232) \pi,\, N(1440)\pi, \, N \rho,\, N \sigma;$ \
$N(1535)  \leftrightarrow  N \eta;$
$N N  \leftrightarrow  N N \pi$;
\item meson-meson collisions
$\rho  \leftrightarrow  \pi \pi \,  \mbox{(p-wave)},
\sigma  \leftrightarrow  \pi \pi \, \mbox{(s-wave)}$
\end{itemize}

\subsection{$\pi N$ total reaction cross sections}

In order to describe pion-baryon scattering in the framework of the
resonance picture a Breit-Wigner formulation for the cross sections
(eq. $\pi N \rightarrow m N$) was used
\begin{equation}
\label{breit}
\sigma_{ab \rightarrow R \rightarrow cd} =
\frac{2J_R+1}{(2S_a+1)(2S_b+1)}\, \frac{4 \pi}{p_i^2}
\frac{s\,\Gamma_{R \rightarrow ab}\, \Gamma_{R \rightarrow cd}}
{(s-M_R^2)^2+s\, \Gamma_{tot}^2}.
\end{equation}
In Eq. (\ref{breit}) $ab$ and $cd$ denote the baryon and the meson in
the initial and final state of the reaction and $R$ is the intermediate
baryon resonance. $J_R$, $S_a$ and $S_b$ are the spins of the baryon
resonance and the particles in the initial state of the reaction,
while $\Gamma_{R\to ab}, \ \Gamma_{R\to cd}$ are taken from the
PDG~\cite{pdg} and $p_i^2$ stands for the squared momentum of the
incoming meson in the resonance rest frame.

The solid line in Fig. \ref{piAfig2} shows the total $\pi^--p$-cross
section within the CBUU model in comparison to the experimental data from 
\cite{landolt}.  To calculate this cross section we replace the
partial widths $\Gamma_{R \to cd}$ in Eq. (\ref{breit}) by the total
widths of the baryonic resonances and sum up the contributions from all
resonances incoherently. The dot-dot-dashed, the dot-dashed and the
short-dashed lines in Fig.  \ref{piAfig2} show the contributions from
the $\Delta(1232)$, the $N(1440)$ and the $N(1520)$ separately.  We
supplement the resonances included in Ref.~\cite{TeisZP97} with an
additional effective two-pion production channel (long-dashed line) in
order to reproduce properly the total cross section data at high
$s^{1/2}$ which is quite essential for $\pi$+A reactions in the energy
regime considered. By using alternatively an effective resonance with a
one-pion decay width we have ascertained that our results are
independent of the specific way in which the missing pion strength at
high energies is corrected.

\subsection{Elementary processes for meson production and
meson-baryon interactions}

Because of the small cross sections involved we can treat the
production of vector mesons ($\rho,\omega$) and $\eta$'s
perturbatively.  Since we work within the parallel ensemble algorithm,
each parallel run of the transport calculation can be considered
approximately as an individual reaction event, where binary reactions
in the entrance channel at given invariant energy $\sqrt{s}$ lead to
final states with 2 or 3 particles with a relative weight $W_i$ for
each event $i$. $W_i$ is defined by the ratio of the production cross
section to the total hadron-hadron cross section\footnote{The actual
final states are chosen by Monte Carlo sampling according to the 2 or
3-body phase space.}.  The perturbative treatment implies that the
initial hadrons are not modified in their respective final channels.  On
the other hand, each perturbative particle is represented by a
testparticle with weight $W_i$ and propagated according to the Hamilton
equations of motion. Elastic and inelastic reactions with baryons are
computed in the standard way~\cite{CassMos90}.  The final cross section
is obtained by multiplying each testparticle with its weight $W_i$.  In
this way one achieves a time-saving simulation of the vector meson
production, propagation and reabsorption during the pion-nucleus
collision.

The $\eta$ mesons are produced in pion-baryon and baryon-baryon
collisions according to the elementary production cross sections from
Refs.~\cite{Wolf90,Vetter}. However, in the present analysis we assume the
$pn \to pn\eta$ cross section to be about 6 times larger than the $pp
\to pp\eta$ cross section close to threshold in line with the new data from 
the WASA collaboration~\cite{WASA}.

For the vector mesons ($\rho, \omega$) we have to take into account
the pion-baryon production channels
$\pi^- p\to \omega n, \pi^-N \to \omega X, \pi^- p \to \rho n,
\pi^-N \to \rho X$ and in addition to \cite{CGIK97,GKC97}
the baryon-baryon channels $BB \to \rho BB$, $BB \to \rho X$,
$BB \to \omega BB$, $BB \to \omega X$.

For the exclusive process $\pi^- p \to \omega n$ we use a parametrization
of the experimental data from Ref. \cite{Cugnon}:
\begin{eqnarray}
\sigma^{excl.}_{\pi^- p \to \omega n} = C \frac{p_{\pi N}-p_{\omega}^0}
{p_{\pi N}^{\alpha}-d},
\label{pip}\end{eqnarray}
where $p_{\pi N}$ is the relative momentum (in GeV/c) of the
pion-nucleon pair while $p_{\omega}^0$ = 1.095 GeV/c is the threshold
value. The parameters $C = $ 13.76 mb (GeV/c)$^{\alpha -1}$,
$\alpha$=3.33 and $d$=1.07 (GeV/c)$^{\alpha}$ describe satisfactorily
the data on the energy-dependent cross section in the near-threshold
energy region. For $\rho^0$-meson  production $\pi^- p \to \rho n$ we
use the same cross section as for the $\omega$-meson; this holds
experimentally within 20\%.

For the inclusive vector meson production $\pi^- p \to V X,
\ V=\omega, \rho$  we use the parametrization from Ref. \cite{SibCM97}
\begin{eqnarray}
\sigma^{incl.}_{\pi^- p \to V X} = a_V (x-1)^{b_V} x^{-c_V},
\label{pipX}\end{eqnarray}
where the scaling variable is defined as $x=s/s_{th}$, $s_{th}=(m_N +m_V)^2$.
For $\omega$-production we have  $a_\omega=4.8$~mb, $b_\omega=1.47$,
$c_\omega=1.26$; for $\rho$-production we have $a_\rho=3.6$~mb,
$b_\rho=1.47$, $c_\rho=1.26$ using $m_\rho$= 0.77 GeV in the vacuum case.
In the actual calculation we take the maximum of the parametrizations
(\ref{pipX}) and (\ref{pip}). These cross sections have been shown in Ref.
\cite{SibCM97} to reproduce the available data very well.

For the vector meson production in baryon-baryon channels (this
contribution is quite small for $\pi^-$A reactions) we also use the
parametrization from Ref. \cite{SibCM97}
\begin{eqnarray}
\sigma^{incl.}_{pp \to V X} = a_V (x-1)^b_V x^{-c_V},
\label{ppX}\end{eqnarray}
where the scaling variable is defined again as $x=s/s_{th}$,
$s_{th}=(2m_N +m_V)^2$.  For $\omega$-production we use
$a_\omega=2.2$~mb, $b_\omega=1.47$, $c_\omega=1.1$; for
$\rho$-production we use $a_\rho=2.5$~mb, $b_\rho=1.47$,
$c_\rho=1.11$.

For the interactions of vector mesons with baryons we include the
following channels:
$\omega N \to \omega N, \  \omega N \to \pi N, \  \omega N \to \pi \pi N$;
$\rho N \to \rho N, \rho N \to \pi N, \rho N \to \pi \pi N$.
Since apart for the $\pi N$ final channels
no experimental data are directly available we
adopt the parametrizations from Ref.~\cite{CGIK97,GKC97}.
The total $\omega N$ cross section is described as
\begin{eqnarray}
\sigma_{\omega N}^{tot} (p_{lab}) = A + {B\over p_{lab}}
\label{omN}\end{eqnarray}
with A = 11 mb and B = 9 mb GeV/c.
The elastic $\omega N$ cross section is  parametrized as
\begin{eqnarray}
\sigma^{el}_{\omega N} (p_{lab}) = A {1\over 1 + a p_{lab}}
\label{omNel}\end{eqnarray}
with A = 20 mb and $a$ = 1 GeV$^{-1}$c.

For the $\rho$-N total and elastic cross sections we adopt the results
of Ref. \cite{Sibirtsev}, which were calculated within the
resonance model using the experimental branching ratios
for the resonances involved \cite{pdg}. A good fit to the results of
Ref. \cite{Sibirtsev} in the energy range of interest is given by:
\begin{eqnarray}
&&\sigma_{\rho^0 N}^{tot} = 26.0 + 0.9 \ p^{-6} \  {\rm [mb]}\nonumber\\
&&\sigma_{\rho^0 N}^{el} = 13.0 + 0.25 \ p^{-6} \  {\rm [mb]},
\label{sib_pr}\end{eqnarray}
where $p$ [GeV/c] is the meson momentum in the cms. Since this
parametrization diverges for zero momentum, we numerically include an
upper limit of 200 mb for the total cross section.

The channel $\rho N \to \pi N$ is determined via
detailed balance from the inverse reaction (\ref{pip}) whereas the
$\omega$ absorption channel is described by $\sigma^{abs}_{\omega N} =
\sigma^{tot}_{\omega N} - \sigma^{el}_{\omega N}$.

In order to demonstrate the relevant range of the elementary production
cross sections we display in Fig.~\ref{piAfig3} the distribution in
the pion-baryon collision number versus the invariant energy $\sqrt{s}$
above the threshold for $\rho$ production $\sqrt{s_{th}} = m_N + m_\rho$,
i.e. $dN/d\sqrt{s}$ (histogram) for $\pi^-$Pb at $E_{kin}=1.3$~GeV.
The arrow indicates the incoming energy.  Due to Fermi motion and
secondary interactions the $dN/d\sqrt{s}$ is not a sharp peak, but a
broader distribution. At the pion kinetic energy of 1.3 GeV one thus is
sensitive to the elementary vector meson
production cross section for excess energies of 100 -- 300 MeV.

We finally note that all the vector mesons are produced and propagated
with their pole mass. Their spectral functions are taken into account
only in their decay to $e^+e^-$. However, the essential broadening of
the $\rho$ spectral function in the nuclear medium due to elastic
and inelastic scattering with nucleons is taken into account dynamically,
though not consistently with the model from Ref.~\cite{Peters}.

\section{Dilepton production}

\subsection{Elementary channels}

The dilepton production is calculated perturbatively by including the
contributions from the Dalitz-decays $\Delta\to N l^+l^-$,
$N(1440)\to N l^+l^-$, $N(1520)\to N l^+l^-$, $N(1535)\to N l^+l^-$,
$\pi^0 \to \gamma l^+l^-$, $\eta \to \gamma l^+l^-$, $\omega
\to \pi^0 l^+l^-$ and the direct dilepton decays of the vector mesons
$\rho$ and $\omega$.
For a detailed description of the $\Delta, N^*$
Dalitz decays we refer the reader to  Ref.~\cite{Wolf90}; the dilepton
decays of the $\eta$ and vector mesons are described in Ref.~\cite{Brat97}.

The novel channel included as compared to Refs.~\cite{CGIK97,GKC97} is
the Dalitz decay of the $N(1520)$ resonance which is described in the
same way as the $\Delta$ resonance with spin 3/2 (see \cite{Wolf90}),
but using a coupling constant $g=0.96$ and $\Gamma_0 = 1.1$~MeV in Eqs.
(4.8)-(4.13) in Ref.~\cite{Wolf90}.
We include this resonance because its presence was recently shown to
dominate the $\rho$-meson properties in a baryon-rich environment
\cite{Peters}.
We also note, that compared to the INC calculations in Refs.
\cite{CGIK97,GKC97} we employ different parametrizations for the
$\eta$-production channels based on the more recent data from the WASA
and PINOT Collaborations \cite{WASA,Pinot}.

The dilepton radiation resulting from $\pi N$ interactions is dominantly
made up by two contributions. One consists of reactions in which the pion
is absorbed on the nucleon to form a nucleon resonance which then later
undergoes a Dalitz decay into $Ne^+e^-$; these processes are explicitly
contained in the transport calculations.  The second class of reactions
consists of those in which the pion reemerges, so that there may be
bremsstrahlung from the external legs of the charged particles.  In
Refs.~\cite{CGIK97,GKC97} a phase-space corrected soft photon
approximation \cite{GaleKap} for the latter processes has been adopted.
However, this approximation has recently been discussed in detail by
Lichard \cite{Lichard} and it has been shown for a particular example
that various approximations, including the one used in
Refs.~\cite{CGIK97,GKC97}, can lead to large uncertainties of up to a
factor of about 5.  This uncertainty estimate agrees with that obtained from 
a comparison with an 'exact' calculation from $pn$ bremsstrahlung
in Fig. 5 in Ref.~\cite{Wolf90}. Since the phase-space corrected soft photon
approximation more likely provides an upper limit for the radiation from
the external legs \cite{Lichard,Eggers} and the $\pi N$
bremsstrahlung channel was found in Ref.~\cite{GKC97} to be a minor
background in the vector-meson mass regime, we discard an explicit
calculation of this channel in this work. We note, however, that this
bremsstrahlung contribution might be of similar order as the Dalitz
decay of the $\eta$-meson at lower invariant masses \cite{GKC97}.
Thus by measuring the $\eta$-yield from its $2\gamma$-decay independently
one might subtract the $\eta$ Dalitz channel from the dilepton mass
spectrum -- as well as for the $\pi^0$, respectively  -- and obtain
experimental bounds on this channel, too.

\subsection{In-medium vector mesons and collisional broadening}

\subsubsection{Collisional broadening}
The production and propagation of short lived hadronic resonances with
all their off-shell properties in the nuclear medium is presently an
unsolved problem and especially the production of $\rho$-mesons close
to threshold is uncertain since its width changes drastically in the
nucleus as compared to the vacuum. We thus have to introduce a couple of
simplifying assumptions for this explorative study that finally have to
be controlled by experimental data.
Following \cite{CGIK97,GKC97} we assume that (in first order) the
in-medium resonance can also be described by a Breit-Wigner formula
with a mass and width distorted by the nuclear environment,
\begin{eqnarray}
\label{BW}
F(M)= \frac{1}{2 \pi} \frac{\Gamma^*_V}{(M-m^*_V)^2 + \Gamma^{*2}_V/4} ,
\end{eqnarray}
containing the effects of collisional broadening,
\begin{eqnarray}
\label{gammas}
\Gamma^*_V=\Gamma_V+\delta \Gamma ,
\end{eqnarray}
where
\begin{eqnarray}
\label{dgamma}
\delta\Gamma = \gamma v\sigma_{VN}^{tot}\rho_N,
\end{eqnarray}
as well as a shift of the meson mass
\begin{eqnarray}
\label{mstar}
m^{*}_R = m_V+\delta m_V.
\end{eqnarray}
In Eq.~(\ref{dgamma}) $v$ is the resonance velocity with respect to the
target at rest, $\gamma$ is the associated Lorentz factor, $\rho_N$ is
the nuclear density and $\sigma_{VN}^{tot}$ is the meson-nucleon total
cross section. For $\sigma_{\rho N}^{tot}$ we use
Eq.~(\ref{sib_pr}) while for $\sigma_{\omega N}^{tot}$ we adopt
Eq.~(\ref{omN}). In using (\ref{gammas}) we neglect the decrease in
the width due to the lowered mass of the vector mesons which is small
compared to the collisional broadening.

In Fig.~\ref{piAfig4} we show the width of $\rho$ and $\omega$ mesons
calculated dynamically according to Eqs.~(\ref{gammas}) and (\ref{dgamma})
(open squares) for $\pi^-$Pb at $E_{kin}=1.3$~GeV. The solid lines indicate
a linear fit of the form
\begin{eqnarray}
\Gamma_V^* (\rho_N) = \Gamma_V^0 (\rho_N) \left(1 +
\beta {\rho_N \over \rho_0}\right),
\label{gamv}\end{eqnarray}
where $\beta = 1.55$ for the $\rho$-meson and $\beta=9$ for the
$\omega$-meson.
While the collisional broadening is roughly twice that of the
bare $\rho$-width, it increases the width of the $\omega$-meson by
about one order of magnitude. The latter values imply that the lifetime
of the $\rho$ at density $\rho_0$ drops to $\approx$ 0.7 fm/c while the
lifetime of the $\omega$ meson decreases to $\approx$ 2.7 fm/c at $\rho_0$.
Note that these values agree approximately with those obtained in a
recent refined hadronic model \cite{Klingl96} .

\subsubsection{'Dropping' vector meson mass}

In order to explore the observable consequences of vector meson mass
shifts at finite nuclear density the in-medium vector meson masses are
modelled according to Hatsuda and Lee \cite{H&L92} or Brown/Rho scaling
\cite{BrownRho} as
\begin{eqnarray}
\label{Brown}
m^*_V = m_V(1 - \alpha \rho_N(r) / \rho_0),
\end{eqnarray}
where $\rho_N (r)$ is the nuclear density at the resonance decay,
$\rho_0 = 0.16 fm^{-3}$ and $\alpha \simeq 0.18$ for the $\rho$ and
$\omega$. The latter value of $\alpha$ has also been used in Refs.
\cite{Cass95C,Cass96H} and led to a good description of the dilepton
spectra from the CERES and HELIOS-3 Collaborations. Furthermore, in
Ref.  \cite{Klingl96} a shift of the $\omega$ pole by $\approx$ 120 MeV
is reported, however, no essential shift of the $\rho$-meson pole is
extracted from their dynamical calculations (cf. also
Ref.~\cite{Leonid}).  Thus the parameter $\alpha$ in (\ref{Brown}) has
to be taken with same care and finally to be determined by experiment.

\subsubsection{The $\rho$ spectral function}

While the dropping mass scenario, together with the collisional
broadening, reproduces, at least qualitatively, the results of the more
refined model of Ref.~\cite{Klingl96}, it is an oversimplified
approximation for the $\rho$-meson.
For the $\rho$-meson we, therefore, also include alternatively
the calculated spectral function from Ref. \cite{Peters}.
The implementation of the $\rho$ spectral function into the transport
approach for the calculation of the  dilepton yield from $\rho^0$ decay is
described in Refs.~\cite{CBRW97,BCRW97}. Here we
adopt the same strategy, i.e. the dilepton radiation from $\rho$ mesons
is calculated as
\begin{equation}
\label{SFrho}
{dN_{\rho\to l^+l^-}\over dM} = - Br(M) {2M\over \pi} \
{\rm Im}D_\rho (q_0, q; \rho_N),
\end{equation}
where $D_\rho$ is the $\rho$-meson propagator \cite{Peters} in the
hadronic medium depending on the baryon density $\rho_N$ as well as on
energy $q_0$ and 3-momentum $q\equiv |{\bf q}|$ in the local rest frame
of the baryon current ('comoving' frame).  The invariant mass $M$ is
related to the $\rho$-meson 4-momentum in the nuclear medium as $M^2 =
q_0^2 - q^2$, while ${\rm Im} D_\rho$ is spin averaged over the
longitudinal and transverse part of the $\rho$-propagator
\cite{Peters}.  Furthermore, $Br(M)$ is the branching ratio of the
$\rho$-meson resonance to dileptons which is in principle an explicit
function of the invariant mass M. Vector-Meson-Dominance can not be used
to evaluate this function where a large part of the $\rho$-strength
resides in nucleon resonance-hole excitations \cite{Peters}. For the nucleon
resonances the simple VDM is known to be quite inaccurate \cite{Friman}.
In the absence of any detailed study of this problem, we have used a
constant branching ratio fixed at the resonance mass for simplicity
as in \cite{CBRW97,BCRW97}; in future this simplifying assumption will
have to be improved.

\section{Dileptons from $\pi$A reactions}

We now come to the results of our numerical simulations.
In  Fig.~\ref{piAfig5} we present the calculated dilepton invariant
mass spectra $d\sigma/dM$ for $\pi^-$Pb at the bombarding energy of
$E_{kin}=1.3$~GeV including a finite mass resolution of 10 MeV.
The upper part corresponds to the result calculated
without collisional broadening and free meson masses. The middle part
shows the dilepton spectra calculated including the collisional
broadening effect and 'dropping' masses of $\rho$ and $\omega$ mesons
(Eq.~(\ref{Brown})), whereas the lower part corresponds to calculations
with collisional broadening and dropping mass shift for the $\omega$-meson
and the momentum- and density-dependent spectral function \cite{Peters}
(Eq.~(\ref{SFrho})) for the $\rho$-meson.

The thin lines indicate the individual contributions from the different
production channels; {\it i.e.}~ starting from low $M$: Dalitz decay
$\eta \to \gamma e^+ e^-$ (dotted line), $\Delta \to N e^+ e^-$
(short-dotted line), $\omega \to \pi^0 e^+ e^-$ (dot-dashed line),
$N(1520) \to N e^+ e^-$ (dot-dashed line), $N(1535) \to N e^+ e^-$
(long-dashed line); for $M \approx $ 0.8 GeV:
$\omega \to e^+e^-$ (dashed line), $\rho^0 \to e^+e^-$ (dot-dashed line).
The full solid line represents the sum of all sources considered here.

The dominant background processes in the low mass region up to
$M\simeq 500$~MeV are the $\eta$ and $\omega$ Dalitz decays
and possibly $\pi N$ bremsstrahlung, which is discarded here.
The contributions from heavy-baryon  resonances are negligibly small.
Above $M\sim 0.6$~GeV the spectrum is dominated by the vector meson
decays with a low background from other hadronic sources.
Our results are in qualitative agreement with the INC calculations in
Refs.  \cite{CGIK97,GKC97} but differ in the individual channels by up
to factors of 2 which is due to the use of different elementary cross
sections as mentioned already at the end of section A and to the
explicit treatment of isospin in our transport approach.

As seen from the middle part, the mass shift and the collisional
broadening effect leads to a two-component structure for the $\omega$
meson:  the narrow peak at $M=m_\omega$ comes from the $\omega$ decaying
outside the nucleus whereas the broader peak corresponds to the decay
inside the nucleus. For the $\rho$ meson this effect is not seen
because the $\rho$ mesons practically all decay inside the nucleus due to
their short lifetime; only the width of the $\rho$ meson peak
becomes larger due to collisional broadening.  The 'dropping' mass for
the $\rho$ and $\omega$ mesons leads to an essential reduction of the
vector meson production threshold in meson-baryon and baryon-baryon
collisions and to a slight enhancement of vector meson production in
pion-nucleus collisions.  A similar effect is seen if one employs the
$\rho$ spectral function (lower part in Fig.~\ref{piAfig5}) instead of
a dropping $\rho$-mass.

We have also performed calculations for light systems (in order
to test different density regimes) and two incoming energies.  In
Figs.~\ref{piAfig6} and \ref{piAfig7} we show the dilepton invariant mass
spectra $d\sigma/dM$ for $\pi^-$C, $\pi^-$Ca and $\pi^-$Pb at the
bombarding energies of $E_{kin}=1.3$ and 1.7~GeV, respectively.  The
solid lines indicate the sum of all dilepton channels calculated
without collisional broadening and with 'free' meson masses.  The dotted
lines are the results with collisional broadening and a 'dropping' mass
of $\rho$ and $\omega$ mesons.  The dashed lines correspond to
calculations with collisional broadening, with dropping $\omega$ mass
and with the $\rho$ spectral function \cite{Peters}.  The 'dropping mass'
scheme leads to an enhancement by about a factor 2 as compared to the
free meson mass in the dilepton range $0.6\le M \le 0.7$~GeV and to a
reduction by about a factor of 2 above the $\omega$ peak.  The $\rho$
spectral function also slightly enhances the yield at $0.6\le M \le
0.7$~GeV, however, the reduction above the $\omega$ peak is not so
strong as compared to the 'free' meson mass case.  As seen from
Figs.~\ref{piAfig6} and \ref{piAfig7} the difference between the three
scenarios exists even for a light system like $^{12}$C and becomes more
pronounced for heavy nuclei like $^{208}$Pb due to the larger volume.

In Fig.~\ref{piAfig8} we present the same analysis as in
Fig.~\ref{piAfig7}, but the dilepton spectra were calculated including a
cut in the longitudinal momentum of the dileptons, i.e. $q_z \le 0.3$~GeV/c,
which makes the differences between the 'dropping mass' scheme and the $\rho$
spectral function more pronounced.  Fig.~\ref{piAfig8} demonstrates that
with a proper $q_z$ cut one should be able to distinguish experimentally
between the two scenarios for the medium modifications of the vector mesons.

\section{Summary}

On the basis of the Coupled-Channel-BUU approach (CBUU) \cite{TeisZP97} we
have studied dilepton production in $\pi^-$C, $\pi^-$Ca and $\pi^-$Pb
collisions at $E_{kin}=1.3$ and 1.7~GeV.  Various contributions are
taken into account for dilepton production:  the Dalitz-decays of
$\Delta, N(1440), N(1520), N(1535)$ resonances and $\eta, \omega$
mesons as well as the direct dilepton decays of the vector mesons
$\rho$ and $\omega$.

The contributions from vector mesons were calculated including the
collisional broadening  and by applying two different in-medium
modification schemes of the vector mesons: i) the dropping mass and
ii) the hadronic spectral function approach \cite{Peters}.

It was found that both scenarios lead to
an enhancement of the dilepton yield at $0.6 \le M \le 0.77$~GeV.
However, the models predict quite different absolute values for the
dilepton yield at invariant masses mainly below and above $M=m_V$
with the total in-medium effect  amounting up to a factor of 3.
This sensitivity, which is comparable to that achieved in
heavy-ion reactions \cite{CBRW97},  can be studied experimentally especially
after applying  proper longitudinal momentum cuts and comparing directly
the spectra from light and heavy targets such as $^{12}$C or $^{208}$Pb.

\acknowledgements
The authors are grateful for various discussions with H. Bokemeyer,
Ye.S. Golubeva, W. Koenig, L.A. Kondratyuk, W. K\"uhn, V.~Metag, W.
Sch\"on and A.~Sibirtsev. Furthermore, they are indebted to M. Post and
W. Peters for the $\rho$ spectral function from Ref.~\cite{Peters}.


\newpage

\begin{figure}[h]
\caption{The binding energy per nucleon as a function of the nucleus
mass number.  The solid curve corresponds to the binding energy
calculated with the symmetry potential $U^{symm}$ whereas the dashed
curve is the result of a calculation without $U^{symm}$. The dotted
line corresponds to the empirical values according to
the liquid drop formula.}
\label{piAfig1}
\end{figure}

\begin{figure}[h]
\caption{The calculated total $\pi^--p$-cross section in comparison to the
data from \protect\cite{landolt} (solid line).
The dot-dot-dashed, the dot-dashed and the short-dashed
lines show the contributions from the $\Delta(1232)$, the $N(1440)$
and the $N(1520)$ separately. The long-dashed line indicates the
contribution of the additional two pion production channel.}
\label{piAfig2}
\end{figure}

\begin{figure}[h]
\caption{The distribution in the pion-baryon collision number versus
the invariant energy $\sqrt{s}$ above the threshold for $\rho$ production
$\sqrt{s_0}=m_N+m_\rho$, i.e.  $dN/d\sqrt{s}$ (histogram), for $\pi^-$Pb
at $E_{kin}=1.3$~GeV.  The arrow indicates the incomming energy.}
\label{piAfig3}
\end{figure}

\begin{figure}[h]
\caption{ The width of the $\rho$ and $\omega$ mesons calculated according
to Eqs.~(\protect\ref{gammas}),(\protect\ref{dgamma}) (open
squares) for $\pi^-$Pb at $E_{kin}=1.3$~GeV. The solid lines
indicate a linear fit with density according to Eq.~(\protect\ref{gamv}).}
\label{piAfig4}
\end{figure}

\begin{figure}[h]
\caption{
The dilepton invariant mass spectra $d\sigma/dM$ for $\pi^-$Pb at the
bombarding energy of $E_{kin}=1.3$~GeV calculated without collisional
broadening and with free meson masses (upper part), including the collisional
broadening effect and a 'dropping' mass of the $\rho$ and $\omega$ mesons
(middle part) as well as including the $\rho$ spectral function from
Ref.~\protect\cite{Peters} instead of a 'dropping' $\rho$-mass (lower part).
The thin lines indicate the individual contributions from the different
production channels; {\it i.e.}~ starting from low $M$: Dalitz decay
$\eta \to \gamma e^+ e^-$ (dotted line), $\Delta \to N e^+ e^-$
(short-dotted line), $\omega \to \pi^0 e^+ e^-$ (dot-dashed line),
$N(1520) \to N e^+ e^-$ (dot-dashed line), $N(1535) \to N e^+ e^-$
(long-dashed line); for $M \approx $ 0.8 GeV:
$\omega \to e^+e^-$ (dashed line), $\rho^0 \to e^+e^-$ (dot-dashed line).
The full solid line represents the sum of all sources. }
\label{piAfig5}
\end{figure}

\begin{figure}[h]
\caption{The dilepton invariant mass spectra $d\sigma/dM$ for $\pi^-
C$, $\pi^-$Ca and $\pi^-$Pb at the bombarding energy of
$E_{kin}=1.3$~GeV.  The solid lines indicate the sum of all dilepton
channel calculated without collisional broadening and with free meson
masses.  The dotted curves are the result with collisional broadening
and 'dropping' masses of $\rho$ and $\omega$ mesons.  The dashed curves
coresspond to the calculations with collisional broadening,
dropping $\omega$ mass and with the $\rho$ spectral function from
Ref.~\protect\cite{Peters}. }
\label{piAfig6}
\end{figure}

\begin{figure}[h]
\caption{The dilepton invariant mass spectra $d\sigma/dM$ for
$\pi^-$C, $\pi^-$Ca and $\pi^-$Pb at the bombarding energy of
$E_{kin}=1.7$~GeV. The assignment is the same as in
Fig.~\protect\ref{piAfig6}.}
\label{piAfig7}
\end{figure}

\begin{figure}[h]
\caption{The dilepton invariant mass spectra $d\sigma/dM$ for
$\pi^-$C, $\pi^-$Ca and $\pi^-$Pb at the bombarding energy of
$E_{kin}=1.7$~GeV with a longitudinal momentum cut $q_z \le 0.3$~GeV/c.
The assignment is the same as in  Fig.~\protect\ref{piAfig6}.}
\label{piAfig8}
\end{figure}

\newpage
\psfig{figure=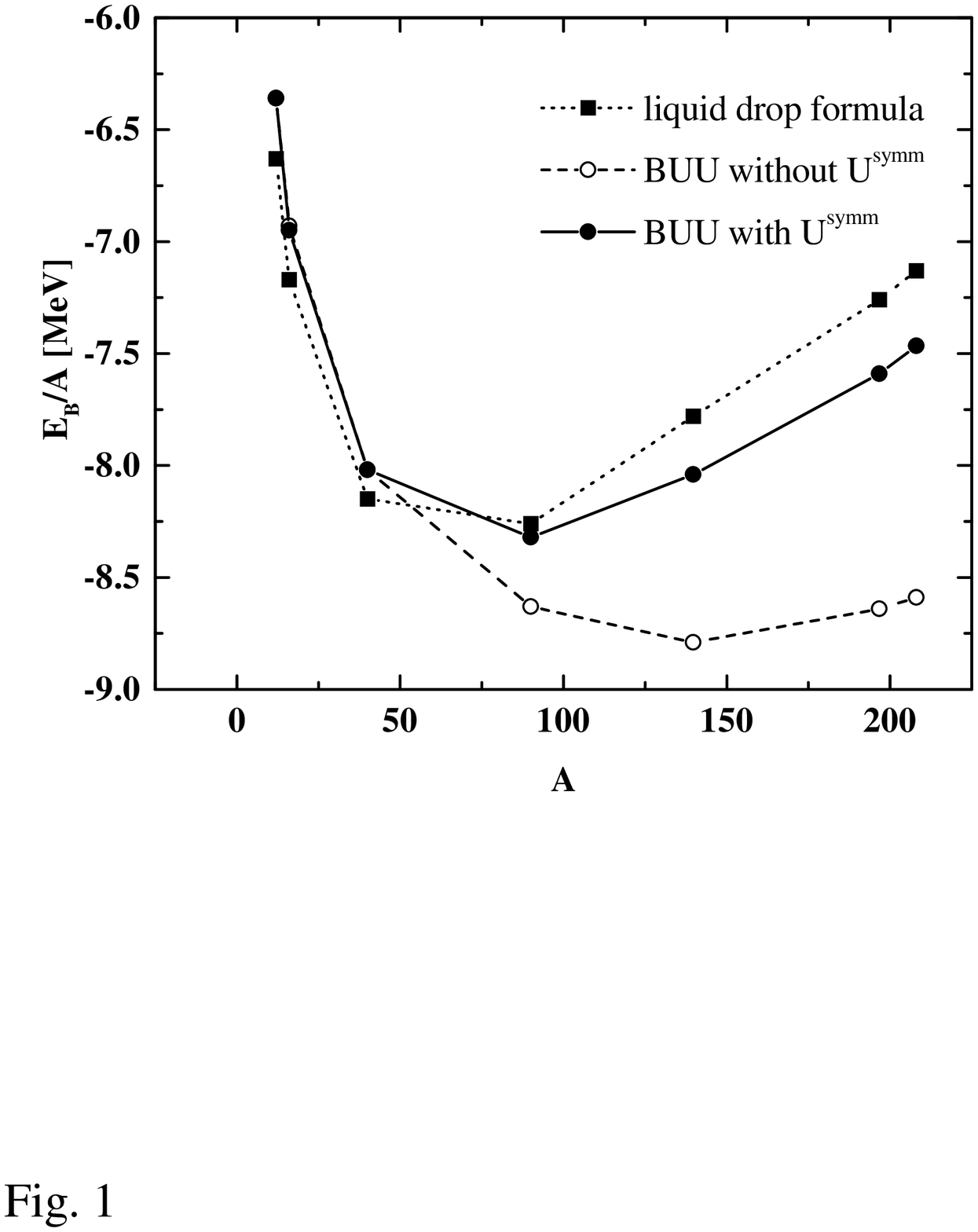,width=15cm,height=22cm}
\psfig{figure=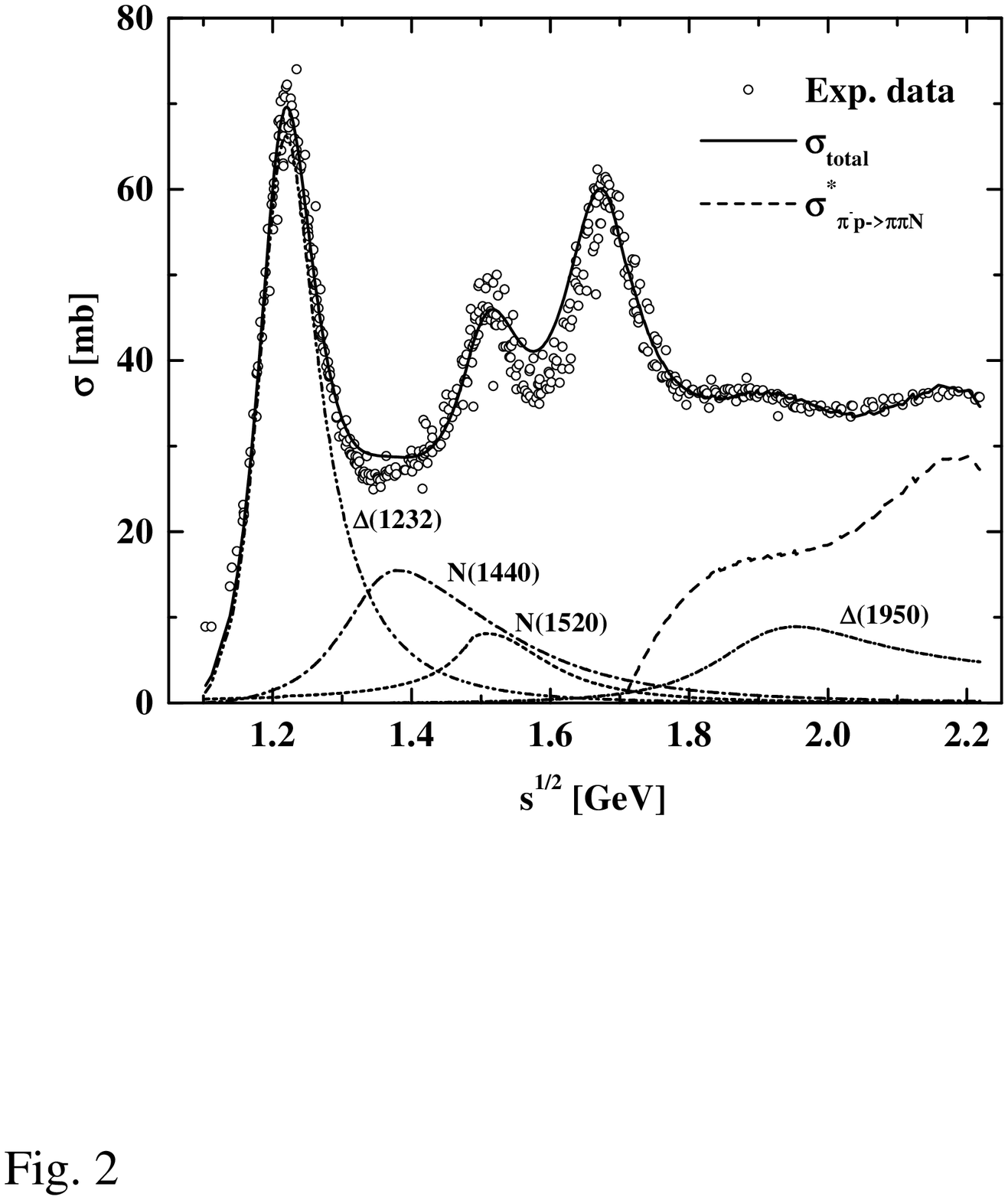,width=15cm,height=22cm}
\psfig{figure=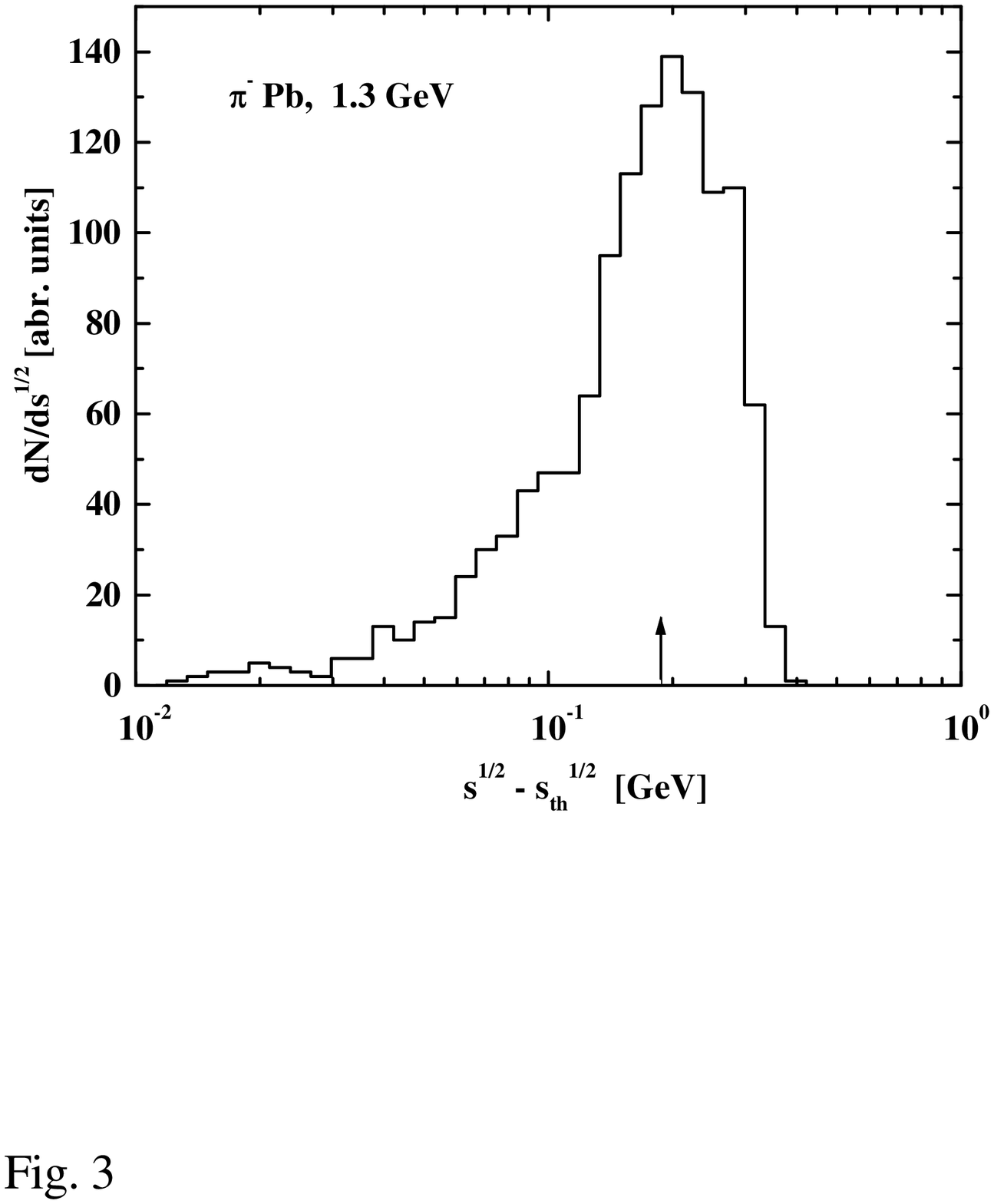,width=15cm,height=22cm}
\psfig{figure=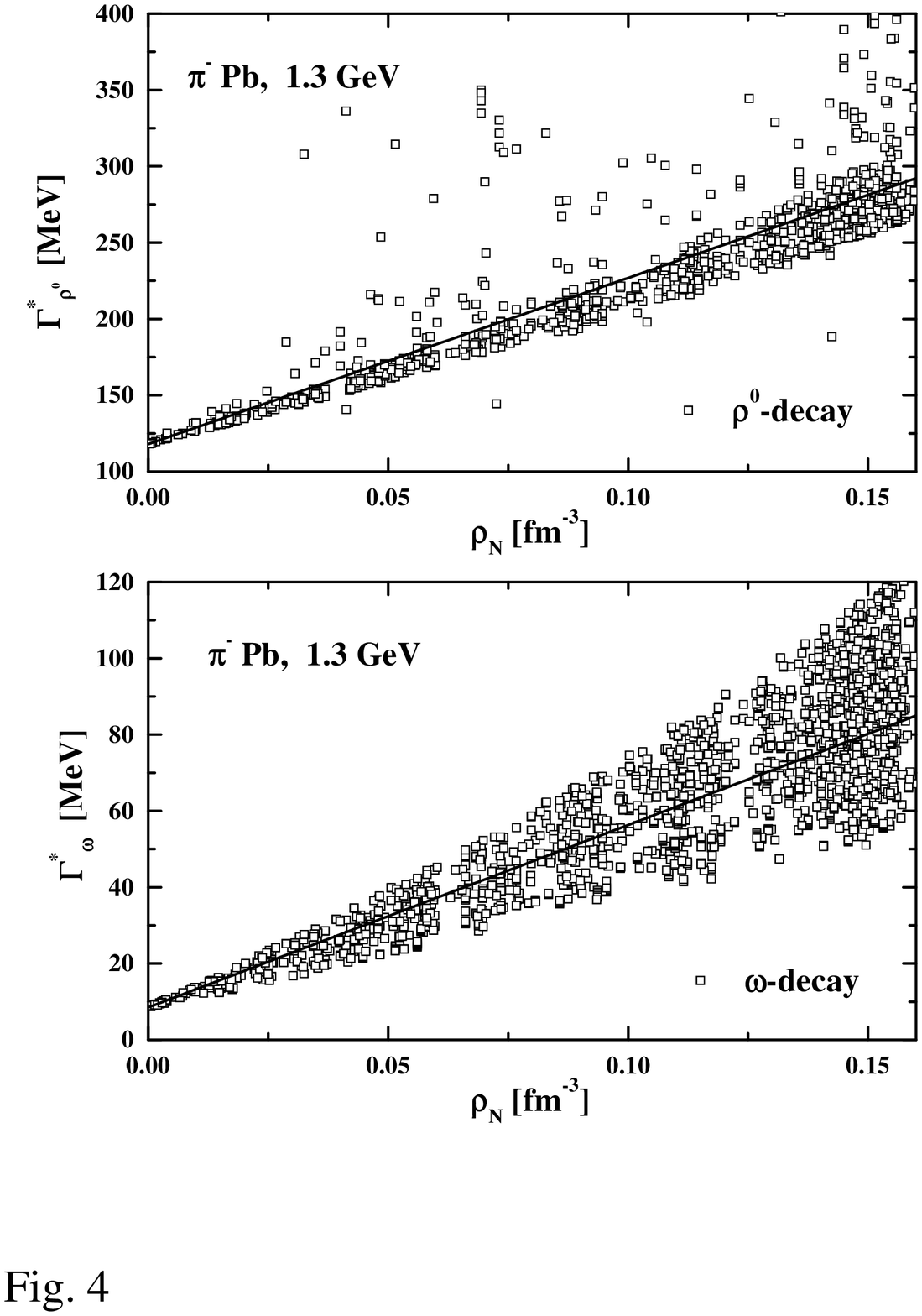,width=15cm,height=22cm}
\psfig{figure=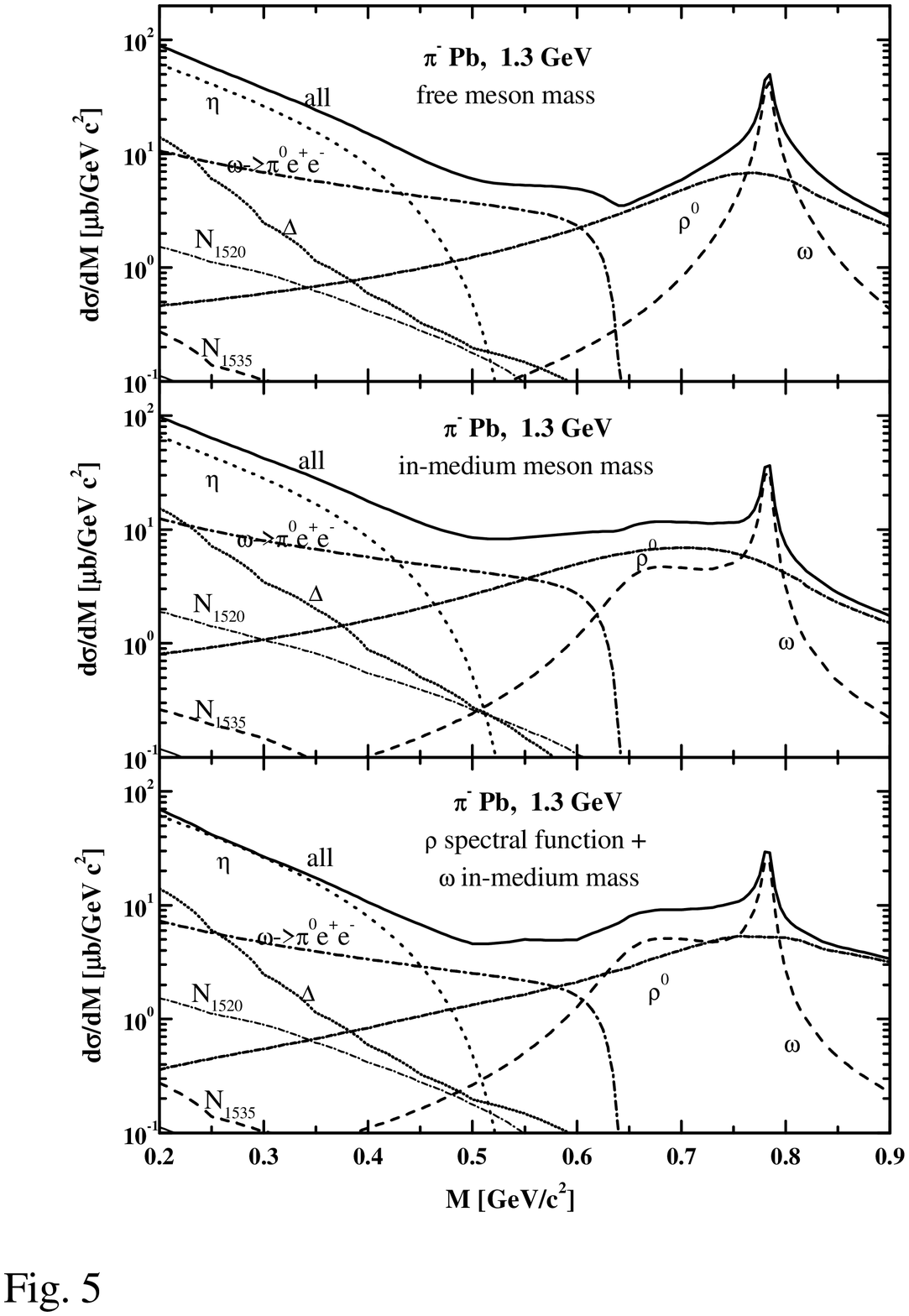,width=15cm,height=22cm}
\psfig{figure=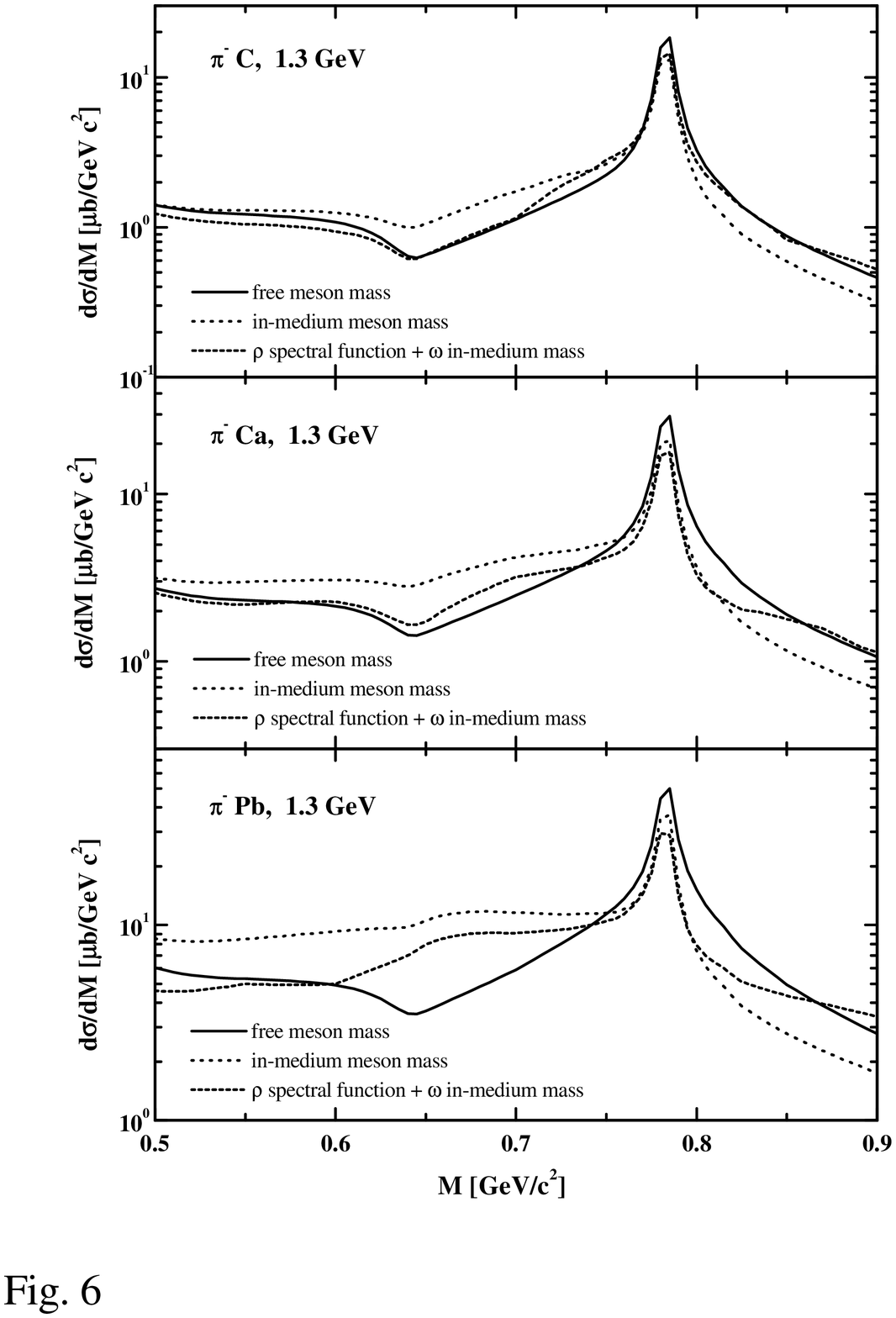,width=15cm,height=22cm}
\psfig{figure=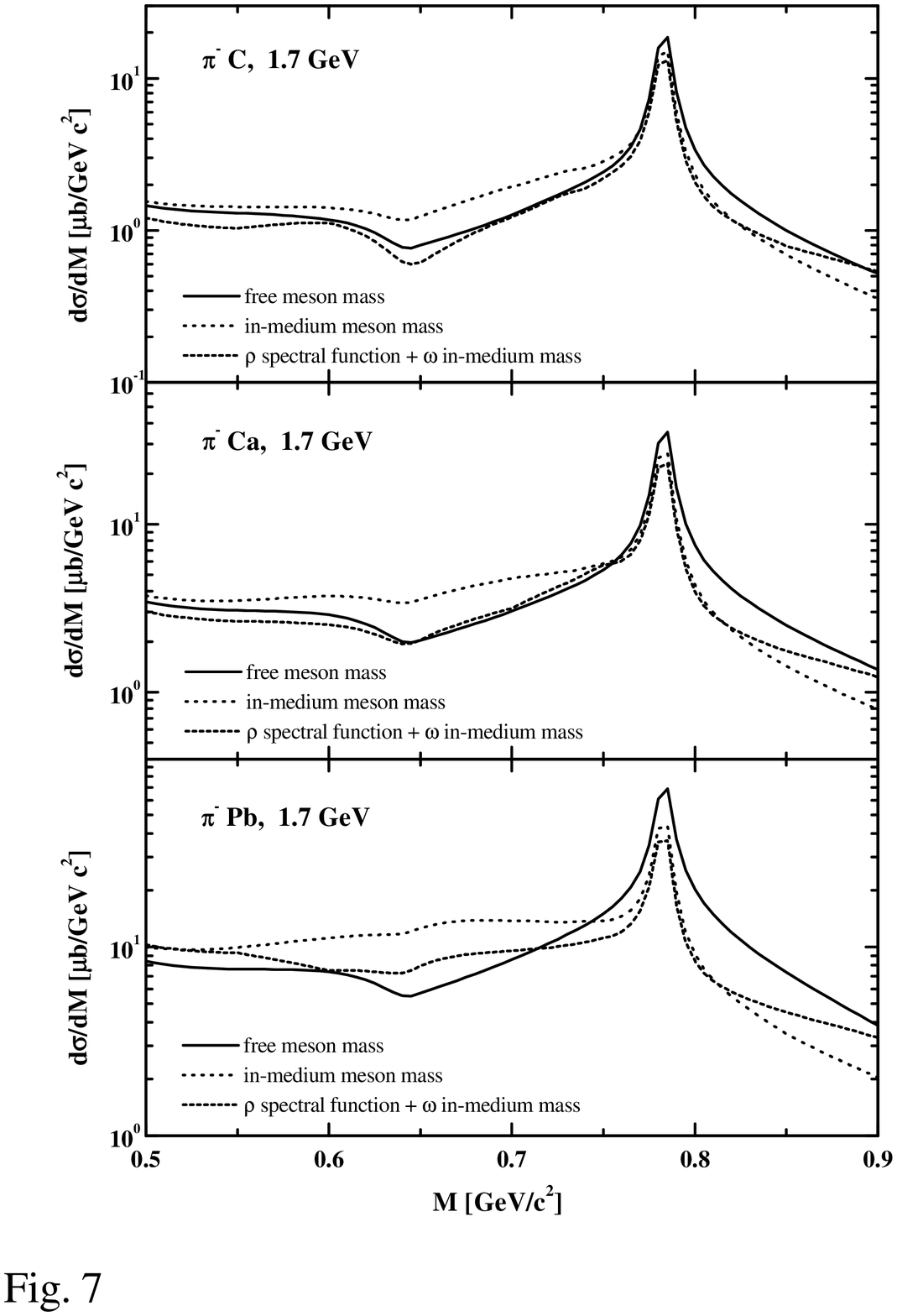,width=15cm,height=22cm}
\psfig{figure=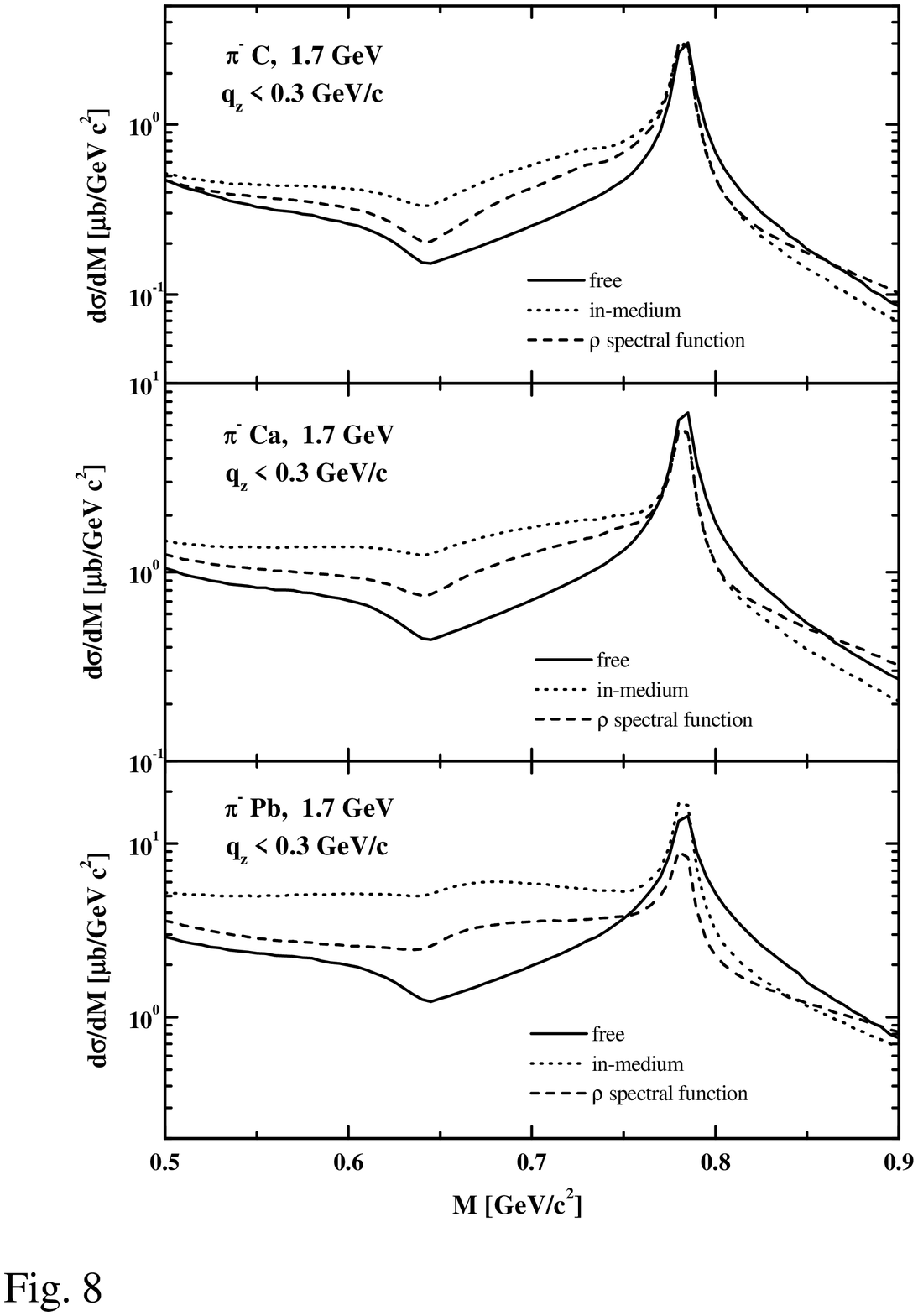,width=15cm,height=22cm}

\end{document}